# A new perspective on steady-state cosmology: from Einstein to Hoyle

Cormac O'Raifeartaigh [ϕ] and Simon Mitton [ϕϕ]

[ϕ]*School of Science, Waterford Institute of Technology, Cork Road, Waterford, Ireland*
[ϕϕ]*Department of History and Philosophy of Science, University of Cambridge, Cambridge, United Kingdom*

Author for correspondence: coraifeartaigh@wit.ie

## Abstract

We recently reported the discovery of an unpublished manuscript by Albert Einstein in which he attempted a 'steady-state' model of the universe, i.e., a cosmic model in which the expanding universe remains essentially unchanged due to a continuous formation of matter from empty space. The manuscript was apparently written in early 1931, many years before the steady-state models of Fred Hoyle, Hermann Bondi and Thomas Gold. We compare Einstein's steady-state cosmology with that of Hoyle, Bondi and Gold and consider the reasons Einstein abandoned his model. The relevance of steady-state models for today's cosmology is briefly reviewed.

## 1. Introduction

It has recently been discovered that Einstein once explored a 'steady-state' model of the cosmos (O'Raifeartaigh et al. 2014; O'Raifeartaigh 2014; Nussbaumer 2014a). An unpublished manuscript on the Albert Einstein Online Archive (Einstein 1931a) demonstrates that Einstein considered the possibility of a universe that expands but remains essentially unchanged due to a continuous formation of matter from empty space (figure 1).[1] Several aspects of the manuscript indicate that it was written in the early months of 1931, during Einstein's first trip to California.[2] Thus, the paper likely represents Einstein's first attempt at a cosmic model in the wake of emerging evidence for an expanding universe. It appears that he abandoned the idea when he realised that the specific steady-state theory he attempted led to a null solution, as described below.

Many years later, steady-state models of the expanding cosmos were independently proposed by Fred Hoyle, Hermann Bondi and Thomas Gold (Hoyle 1948; Bondi and Gold 1948). The hypothesis formed a well-known alternative to 'big bang' cosmology for some years (Kragh 1996 pp. 186-218; Kragh 2007 pp. 187-206; Nussbaumer and Bieri 2009 pp. 161-163), although it was eventually ruled out by astronomical observation.[3] While it could be argued that steady-state cosmologies are of little practical interest today, we find it most interesting that Einstein conducted an internal debate between steady-state and evolving models of the cosmos decades before a similar debate engulfed the cosmological community. In particular, the episode casts new light on Einstein's journey from a static, bounded cosmology to the dynamic, evolving universe.

## 2. Historical context

Soon after the successful formulation of the general theory of relativity (Einstein 1916), Einstein applied his new theory of gravity, space and time to the universe as a whole.[4]

---

[1] Until now, the manuscript was mistaken for an early draft of Einstein's cosmic model of 1931 (Einstein 1931b). A translation and analysis of the full manuscript can be found in (O'Raifeartaigh et al. 2014).
[2] References to Hubble's observations, a lack of references to Einstein's evolving models of 1931 and 1932, and the fact that the paper is set out on American notepaper make it very likely that the paper was written during Einstein's first visit to Caltech (O'Raifeartaigh et al 2014; Nussbaumer 2014a).
[3] Observations of the distributions of the galaxies at different epochs and the discovery of the cosmic microwave background favoured evolving models of the cosmos. See (Kragh 1996 pp. 318-380) for a review.
[4] A major motivation was the clarification of the conceptual foundations of general relativity, i.e., to establish *"whether the relativity concept can be followed through to the finish, or whether it leads to contradictions"* (Einstein 1917a).

Assuming a cosmos that was static over time,[5] and that a consistent theory of gravitation should incorporate Mach's principle,[6] he found it necessary to add a new term to the general field equations in order to predict a universe with a non-zero mean density of matter - the famous 'cosmological constant' (Einstein 1917b).[7] With judicious choice of the cosmological constant, Einstein was led to a model of a finite, static cosmos of spherical spatial geometry whose radius was directly related to the density of matter.[8]

That same year, the Dutch theorist Willem de Sitter proposed an alternative relativistic model of the cosmos, namely the case of a static universe empty of matter (de Sitter 1917). Einstein was greatly perturbed by de Sitter's solution, as it suggested a spacetime metric that was independent of the matter it contained, in conflict with Einstein's understanding of Mach's principle.[9] The de Sitter model became a source of some confusion amongst theorists for some years; it was later realised that the model was not static (Weyl 1923; Lemaître 1925). However, the solution attracted some attention in the 1920s because it predicted that the radiation emitted by test particles inserted into the 'empty' universe would be red-shifted, a prediction that chimed with emerging astronomical observations.[10]

In 1922, the young Russian physicist Alexander Friedman suggested that non-stationary solutions to the Einstein field equations should be considered in relativistic models of the cosmos (Friedman 1922). With a second paper in 1924, Friedman explored almost all the main theoretical possibilities for the evolution of the cosmos and its geometry (Friedman 1924). However, Einstein did not welcome Friedman's time-varying models of the cosmos. His first reaction was that Friedman had made a mathematical error (Einstein 1922). When Friedman showed that the error lay in Einstein's correction, Einstein duly retracted it (Einstein 1923a); however, an unpublished draft of Einstein's retraction makes it clear that he considered time-varying models of the cosmos to be unrealistic:*"to this a physical significance can hardly be ascribed"* (Einstein 1923b).[11]

---

[5] No evidence to the contrary was known to Einstein at the time.
[6] Einstein's view of Mach's principle in these years was that space could not have an existence independent of matter.
[7] It was also assumed that the universe was homogeneous and isotropic on the largest scales
[8] Einstein's suggestion was that a new term comprising the fundamental tensor $g_{\mu\nu}$ multiplied by a universal constant $\lambda$ could be added to the field equations without destroying the general covariance. This term resulted in a static universe of closed curvature, neatly removing the problem of boundary conditions. However, it was later shown that this solution is unstable against the slightest inhomogeneity in matter (Eddington 1930).
[9] A review of Einstein's objection to the de Sitter universe can be found in (Berstein and Feinberg 1986) pp 10-11, (Earman 2001) and (Nussbaumer and Bieri 2009) p78.
[10] Observations of the redshifts of the spiral nebulae were published by VM Slipher in 1915 and 1917 (Slipher 1915, 1917), and became widely known when they were included in a book on relativity and cosmology by Arthur Eddington (Eddington 1923).

Unaware of Friedman's analysis, the Belgian physicist Georges Lemaître proposed an expanding model of the cosmos in 1927. A theoretician with significant training in astronomy, Lemaître was aware of V.M. Slipher's observations of the redshifts of the spiral nebulae (Slipher 1915, 1917), and of Edwin Hubble's emerging measurements (Hubble 1925) of the vast distances to the nebulae (Kragh 1996 p29; Farrell 2009 p78, p90). Interpreting Slipher's redshifts as a relativistic expansion of space, Lemaître derived a universe of expanding radius from Einstein's field equations, and estimated a rate of cosmic expansion from average values of the velocities and distances of the nebulae from Slipher and Hubble respectively.[12] This work received very little attention at first, probably because it was published in French in a little-known Belgian journal (Lemaître 1927). However, Lemaître discussed the model directly with Einstein at the 1927 Solvay conference, only to have it dismissed with the forthright comment:*"Vos calculs sont corrects, mais votre physique est abominable"* (Lemaître 1958).[13]

In 1929, Edwin Hubble published the first empirical evidence of a linear relation between the redshifts of the spiral nebulae (now known to be extra-galactic) and their radial distance (Hubble 1929).[14] By this stage, it had also been established that the static models of Einstein and de Sitter presented problems of a theoretical nature.[15] In consequence, the notion of a relativistic cosmic expansion was taken seriously,[16] and a variety of time-varying models of the cosmos of the Friedman-Lemaître type were advanced (Eddington 1930, 1931: de Sitter 1930a, 1930b; Tolman 1930a, 1930b, 1931, 1932; Heckmann 1931, 1932; Robertson 1932, 1933).

By 1931, Einstein had accepted the dynamic universe. During a three-month sojourn at Caltech in Pasadena in early 1931, a trip that included a meeting with the astronomers of

---

[11] Einstein wisely withdrew the remark before publication. A more detailed account of this episode can be found in (Nussbaumer and Bieri 2009) pp 91-92 or (Nussbaumer 2014b).
[12] He obtained a value of 625 km s$^{-1}$ Mpc$^{-1}$, in reasonable agreement with that estimated by Hubble two years later.
[13] It was on this occasion that Lemaître first learnt of the earlier work of Alexander Friedman (Lemaître 1958).
[14] It has recently been argued that Hubble's 1929 graph was far from definitive due to a number of misclassifications (Peacock 2013). However, many physicists found the result quite convincing at the time.
[15] Einstein's universe was not stable (Lemaître 1927; Eddington 1930) while de Sitter's universe was not truly static (Weyl 1923; Lemaître 1925).
[16] At a meeting of the Royal Astronomical Society in January 1930, de Sitter noted that static models of the cosmos were not compatible with Hubble's observations. In the ensuing discussion, Eddington suggested that a new model of the cosmos was needed (Nussbaumer and Bieri 2009 p121). Following a communication from Lemaître, Eddington arranged for Lemaître's 1927 paper to be translated into English and published in the Proceedings of the Royal Astronomical Society (Lemaître 1931a).

Mount Wilson Observatory and regular discussions with the Caltech theorist Richard Tolman,[17] Einstein made several public statements to the effect that he viewed Hubble's observations as likely evidence for a cosmic expansion. For example, the *New York Times* reported Einstein as commenting that *"New observations by Hubble and Humason concerning the redshift of light in distant nebulae makes the presumptions near that the general structure of the universe is not static"* (AP 1931a) and *"The redshift of the distant nebulae have smashed my old construction like a hammer blow"* (AP 1931b). Not long afterwards, Einstein published two distinct dynamic models of the cosmos, the Friedman-Einstein model of 1931 and the Einstein-de Sitter model of 1932 (Einstein 1931b; Einstein and de Sitter 1932).

Written in April 1931,[18] the Friedman-Einstein model marked the first scientific publication in which Einstein formally abandoned the static universe. Citing Hubble's observations, Einstein suggested that the assumption of a static universe was no longer justified: *"Now that it has become clear from Hubbel's results that the extra-galactic nebulae are uniformly distributed throughout space and are in dilatory motion (at least if their systematic redshifts are to be interpreted as Doppler effects), assumption (2) concerning the static nature of space has no longer any justification."* (Einstein 1931b).[19] Adopting Friedman's 1922 analysis of a universe of time-varying radius and positive spatial curvature,[20] Einstein also removed the cosmological constant he had introduced in 1917, on the grounds that it was now both unsatisfactory (it gave an unstable solution) and unnecessary: *"Under these circumstances, one must ask whether one can account for the facts without the introduction of the λ-term, which is in any case theoretically unsatisfactory"* (Einstein 1931b). The resulting model predicted a cosmos that would undergo an expansion followed by a contraction, and Einstein made use of Hubble's observations to extract estimates for the current radius of the universe, the mean density of matter and the timespan of the expansion. Noting that the latter estimate was less than the ages of the stars estimated from astrophysics, Einstein attributed the paradox to errors introduced by the simplifying assumptions of the models, notably the assumption of homogeneity: *"The greatest difficulty*

---

[17] An account of Einstein's time in Pasadena can be found in (Nussbaumer and Bieri 2009, pp 144-146), (Bartusiak 2009, pp 251-256) and (Eisinger 2011 pp 110-115). It is possible that the seed for Einstein's conversion was planted during his visit to Eddington in the summer of 1930 (Nussbaumer and Bieri 2009, pp …; Nussbaumer 2014b).
[18] It is known from Einstein's diaries that this work was written in the second week of April 1931 and submitted on April 16th (Nussbaumer 2009 pp 146-147; Eisinger 2011 p120).
[19] We have recently presented a first English translation of this work (O'Raifeartaigh and McCann 2014).
[20] It should be noted that the Friedman models included a cosmological constant, as did the de Sitter model.

*with the whole approach, as is well-known, is that the elapsed time since P = 0 comes out at only about $10^{10}$ years….One can seek to escape this difficulty by noting that the inhomogeneity of the distribution of stellar material makes our approximate treatment illusory*" (Einstein 1931b). [21]

In early 1932, Einstein and Willem de Sitter both spent time at Caltech in Pasadena, and they used the occasion to explore a new dynamic model of the cosmos. This model was based on the realisation that a finite density of matter in a non-static universe does not necessarily demand a curvature of space.[22] Mindful of a lack of empirical evidence for spatial curvature, Einstein and de Sitter set this parameter to zero (Einstein and de Sitter 1932). With both the cosmological constant and spatial curvature removed, the resulting model described a cosmos of Euclidean geometry in which the rate of expansion $h$ was related to the mean density of matter $\rho$ by the simple relation $h^2 = \frac{1}{3}\kappa\rho$, with $\kappa$ as the Einstein constant.[23] Applying Hubble's value of 500 km s$^{-1}$ Mpc$^{-1}$ for the recession rate of the galaxies, the authors calculated a value of $4\times10^{-28}$ gcm$^{-3}$ for the mean density of matter, a value that was not incompatible with estimates from astronomy.

The Einstein-de Sitter model played a significant role in the development of 20$^{th}$ century cosmology (North 1965 p134; Kragh 1996 p35; Nussbaumer and Bieri 2009 p152). One reason was that it marked an important hypothetical case in which the expansion of the universe was precisely balanced by a critical density of matter; a cosmos of lower mass density would be of hyperbolic geometry and expand at an ever-increasing rate, while a cosmos of higher mass density would be of spherical geometry and eventually collapse. Another reason was the model's great simplicity; in the absence of any empirical evidence for spatial curvature or a cosmological constant, there was little reason to turn to more complicated models.[24] While the timespan of the expansion was not considered in the rather terse paper of Einstein and de Sitter (Einstein and de Sitter 1932), this aspect of the model was considered by Einstein a year later (Einstein 1933).[25] Noting again that the time of expansion was less than the estimated ages of the stars, he once again attributed the problem to the simplifying assumptions of the model: *"This time-span works out at approximately*

---

[21] In fact, Einstein overestimated the timespan of the expansion by a factor of ten, but his estimate was still small enough to conflict with estimates of stellar age (O'Raifeartaigh and McCann 2014).
[22] This possibility seems to have been overlooked by Friedman (Friedman 1922, 1924) and was first explored by Otto Heckmann (Heckmann 1931).
[23] The pressure of radiation was also assumed to be zero in the model.
[24] Empirical evidence for a positive cosmological constant did not emerge until 1992, while no evidence for spatial curvature has yet been detected.
[25] We have recently provided a first English translation of this little-known paper (O'Raifeartaigh et al. 2015). Once again, Einstein overestimates the time of the expansion by a factor of ten.

*$10^{10}$ years. Of course, at that time the density will not actually have been infinitely large; Laue has rightly pointed out that our rough approximation, according to which the density ρ is independent of location, breaks down for this time"* (Einstein 1933a).

The Einstein-de Sitter model marked Einstein's last original contribution to cosmology; he did not publish any new cosmic models beyond this point.

3. **A brief tour of Einstein's steady-state model**

As noted elsewhere (O'Raifeartaigh et al. 2014; Nussbaumer 2014), several aspects of Einstein's steady-state manuscript suggest that it was written in early 1931, i.e., before his evolving models of 1931 and 1932 (Einstein 1931b; Einstein-de Sitter model of 1932). Thus, the paper almost certainly represents Einstein's first attempt at a relativistic model of the expanding universe.

The manuscript begins with Einstein recalling the well-known problem of gravitational collapse in a Newtonian universe. This starting point is similar to Einstein's static model of 1917 (Einstein 1917), although he now includes a reference to the work of Hugo Seeliger:[26]

> *"It is well known that the most important fundamental difficulty that comes to light when one enquires how the stellar matter fills up space in very large dimensions is that the laws of gravity are not in general consistent with the hypothesis of a finite mean density of matter. At a time when Newton's theory of gravity was still generally accepted, Seeliger had for this reason modified the Newtonian law by the introduction of a distance function that, for large distances r, diminished considerably faster than $1/r^2$."*

Einstein points out that a similar problem arises in relativistic models of the cosmos, and recalls his introduction of the cosmological constant to the field equations of relativity to render them consistent with a static universe of constant radius and matter density:

> *"This difficulty also arises in the general theory of relativity. However, I have shown that this can be overcome through the introduction of the so-called "λ–term" to the field equations. The field equations can then be written in the form*

$$\left(R_{ik} - \frac{1}{2} g_{ik} R\right) - \lambda g_{ik} = \kappa T_{ik} \qquad (1)$$

$$R_{ik} = \Gamma^{\sigma}_{ik,\sigma} - \Gamma^{\sigma}_{i\sigma,k} - \Gamma^{\sigma}_{i\tau}\Gamma^{\tau}_{k\sigma} + \Gamma^{\sigma}_{ik}\Gamma^{\tau}_{\sigma\tau} \qquad (1a)$$

---

[26] In order to avoid the problem of gravitational collapse in the Newtonian universe, Hugo von Seeliger suggested the introduction of an extra term to Newton's law of gravitation that would be effective only at the largest distances (Seeliger 1895).

> *I showed that these equations can be satisfied by a spherical space of constant radius over time, in which matter has a density ρ that is constant over space and time."*

In the next part of the manuscript, Einstein suggests that this static model now seems unlikely:

> *"It has since transpired that this solution is almost certainly ruled out for the theoretical comprehension of space as it really is"*

We note that Einstein dismisses his static model for two separate reasons. First, he comments on the existence of dynamic solutions and that his static solution was found to be unstable:

> *"On the one hand, it follows from research based on the same equations by [ ] and by Tolman that there also exist spherical solutions with a world radius P that is variable over time, and that my solution is not stable with respect to variations of P over time."*

The blank in the sentence above representing theoreticians other than Tolman who suggested dynamic solutions is puzzling as Einstein was unquestionably aware of the cosmological models of both Friedman and Lemaître.[27] Einstein also neglects to make it clear from what research it follows that his static solution is unstable, although this is very likely a reference to Eddington's paper on the subject (Eddington 1930).

Einstein's second reason for ruling out his former static solution concerns the recent astronomical observations of Edwin Hubble:

> *"On the other hand, Hubbel's* [sic] *exceedingly important investigations have shown that the extragalactic nebulae have the following two properties*
>
> 1) *Within the bounds of observational accuracy they are uniformly distributed in space*
> 2) *They possess a Doppler effect proportional to their distance"*

We note that Hubble's name is misspelt throughout the manuscript, as in the case of the Friedman-Einstein model of 1931 (1931b). This may indicate that Einstein was not fully familiar with Hubble's work, as has been argued elsewhere (Nussbaumer and Bieri 1996 p149; Nussbaumer 2014b). We also note that Einstein uses the term 'Doppler effect' rather than radial velocity, suggesting a qualified acceptance of the redshifts of the nebulae as evidence for a cosmic expansion.[28] Remarking that the dynamic models of de Sitter and

---

[27] Einstein's reaction to each is well-known as discussed in section 2.
[28] Many others shared this equivocation at the time, including Hubble (Hubble 1929, 1936).

Tolman are consistent with Hubble's observations, Einstein then points out that their models predict an age for the universe that is problematic:

> *"De Sitter and Tolman have already shown that there are solutions to equation (1) that can account for these observations. However the difficulty arose that the theory unvaryingly led to a beginning in time approximately $10^{10}$-$10^{11}$ years ago, which for various reasons seemed unacceptable."*

We note that there is again no reference to the evolving models of Friedman or Lemaître (Friedman 1922, 1924; Lemaître 1927).The "various reasons" in the quote is almost certainly a reference to the fact that the estimated timespan of dynamic models was not larger than the ages of stars as estimated from astrophysics (see section 2).

In the third part of the manuscript, Einstein explores an alternative solution to the field equations that could also be compatible with Hubble's observations – namely, an expanding universe in which the density of matter does not change over time:

> *"In what follows, I would like to draw attention to a solution to equation (1) that can account for Hubbel's facts, and in which the density is constant over time. While this solution is included in Tolman's general scheme, it does not appear to have been taken into consideration thus far."*

It is not entirely clear what Einstein means by the reference to "Tolman's general scheme''; it may be a reference to a paper in which Tolman suggested that the cosmic expansion might arise from a continuous transformation of matter into radiation (Tolman 1930a).

Einstein starts his analysis by choosing the metric of flat space expanding exponentially:

> *"I let*
>
> $$ds^2 = -e^{\alpha t}(dx_1^2 + dx_2^2 + dx_3^2) + c^2 dt^2 \;....\quad (2)$$
>
> *This manifold is spatially Euclidean. Measured by this criterion, the distance between two points increase over time as $e^{\frac{\alpha}{2}t}$; one can thus account for Hubbel's Doppler effect by giving the masses (thought of as uniformly distributed) constant co-ordinates over time."*

To modern eyes, equation (2) represents the metric of the de Sitter universe. A similar line element was employed by Tolman in the paper mentioned above (Tolman 1930a) and Einstein's choice of metric may owe something to this. However, it should be noted that the hypothesis of a constant rate of matter creation in any case implies a metric that is spatially flat and exponentially expanding.[29] Einstein first notes that the metric is invariant:

> *"Finally, the metric of this manifold is constant over time. For it is transformed by applying the substitution*
>
> $$t' = t - \frac{\alpha}{2}\tau \quad (\tau = const)$$
>
> $$\frac{x'_1}{x_1} = \frac{x'_2}{x_2} = \frac{x'_3}{x_3} = e^{-\frac{\alpha}{2}t}$$
>
> $$ds^2 = e^{\alpha t'}(dx'^2_1 + dx'^2_2 + dx'^2_3) + c^2 dt'^2$$

We note that there appears to be a sign error in the expansion term of the last equation above, an error that may have led to a miscalculation in the analysis described below.

Assuming a low velocity of masses relative to the co-ordinate system and that the gravitational effects of the radiation pressure are negligible, Einstein constructs the matter-energy tensor:

> *"We ignore the velocities of the masses relative to the co-ordinate system as well as the gravitational effect of the radiation pressure. The matter tensor is then to be expressed in the form*
>
> $$T^{ik} = \rho u^i u^k \quad (u^i = \frac{dx^i}{ds})$$
>
> $$T_{ik} = \rho u^\sigma u^\tau g_{\sigma i} g_{\tau k} \quad (3)$$
>
> where $\quad u^1 = u^2 = u^3 = 0; \, u^4 = \frac{1}{c}$ "

From equations (1) - (3), Einstein derives two simultaneous equations and, eliminating the cosmological constant, deduces a relation between the expansion coefficient $\alpha$ and the matter density $\rho$:

> *" Equations (1) yield:*
>
> $$\frac{-3\cancel{(9)}}{4}\alpha^2 + \lambda c^2 = 0$$
>
> $$\frac{3}{4}\alpha^2 - \lambda c^2 = \kappa\rho c^2$$
>
> or $\quad \alpha^2 = \frac{\kappa c^2}{3}\rho \quad (4)$"

---

[29] A constant rate of matter creation implies spatial flatness because the creation rate is affected by spatial curvature ($k/R^2$) and the radius is not constant. A constant Hubble parameter $\dot{R}/R$ is also implied, from which it follows that the expansion must be exponential. It is very possible that Einstein realised this independently of Tolman.

Thus, Einstein concludes from equation (4) that the density of matter $\rho$ remains constant and is related to the cosmic expansion factor $\alpha$:

> *"The density is therefore constant and determines the expansion apart from its sign."*

However, it should be noted that equation (4) is incorrect, and arose from an incorrect derivation of the coefficient of $\alpha^2$ in the first of the simultaneous equations. Einstein later corrected this coefficient from +9/4 to -3/4 (see figure 2), an amendment that leads to the null solution $\rho = 0$ instead of equation (4).

In the final part of the manuscript, Einstein proposes a mechanism to allow the density of matter remain constant in a universe of expanding radius - namely, the continuous formation of matter from empty space:

> *"If one considers a physically bounded volume, particles of matter will be continually leaving it. For the density to remain constant, new particles of matter must be continually formed in the volume from space."*

This proposal closely anticipates the 'creation field' or 'C-field' of Fred Hoyle. However, Einstein has not introduced a term representing this process into the field equations. Instead he associates the continuous formation of matter with the cosmological constant, commenting that the latter ensures that space is not empty of energy:

> *"The conservation law is preserved in that by setting the λ-term, space itself is not empty of energy; as is well-known its validity is guaranteed by equations (1)."*

Thus, in this model of the cosmos, Einstein proposes that the cosmological constant assigns an energy to empty space that in turn allows the creation of matter. However, the proposal is fundamentally flawed because the lack of a specific term representing matter creation in fact leads to the null solution $\rho = 0$. It is clear from the manuscript that Einstein recognized this problem on revision; it seems that he then abandoned the proposal rather than consider more sophisticated steady-state solutions.

### 4. On later steady-state models of the cosmos

The concept of an expanding universe that remains in a steady-state due to a continuous creation of matter is most strongly associated with the Cambridge physicists Fred Hoyle, Hermann Bondi and Thomas Gold. In the late 1940s, these physicists became concerned with well-known problems associated with evolving models of the expanding cosmos. In particular, they noted that the evolving models predicted a cosmic age that was problematic, and disliked Lemaître's idea of a universe with an explosive beginning (Lemaître 1931b, 1931c).[30] In order to circumvent these, and other problems,[31] the trio explored the idea of an expanding universe that does not evolve over time, i.e., a cosmos in which the mean density of matter is maintained constant by a continuous creation of matter from the vacuum (Hoyle 1948; Bondi and Gold 1948).

In the case of Bondi and Gold, the proposal of a steady-state model followed from their belief in the 'perfect cosmological principle', a principle that posited that the universe should appear essentially the same to all observers in all places *at all times*. For an expanding universe, this principle demanded the postulate of a continuous creation of matter. While the idea bears some similarity to Einstein's steady-state model, it is difficult to compare the models directly because the Bondi-Gold theory was not formulated in the context of general relativity.[32]

On the other hand, Fred Hoyle constructed a steady-state model of the cosmos by means of a simple modification of the Einstein field equations (Hoyle 1948; Mitton 2011, chapter 5). Replacing the cosmological constant with a new 'creation-field' term $C_{ik}$, representing the continuous formation of matter from the vacuum, Hoyle obtained the equation

$$\left(R_{ik} - \frac{1}{2} g_{ik} R\right) - C_{ik} = \kappa T_{ik} \quad (5)$$

The new creation-field term allowed for an unchanging universe but was of importance only on the largest scales, in the same manner as the cosmological constant. We note that the perfect cosmological principle followed as a consequence of Hoyle's model, rather than a starting assumption; we also note that Hoyle proposed a more sophisticated version based on the principle of least action some years later (Hoyle and Narlikar 1962).[33]

As is well known, a significant debate was waged between steady-state and evolving models of the cosmos during the 1950s and 1960s (Kragh 1996, chapter 5; Mitton 2011, chapter 7). Eventually, steady-state models were ruled out by astronomical observations[34] that showed unequivocally that we inhabit a universe that is evolving over time.[35] We note that there is no evidence to suggest that any of the steady-state theorists were aware of the manuscript under discussion; indeed, it is likely that they would have been greatly intrigued to learn that Einstein had once attempted a steady-state model.

---

[30] It should be noted that the problem of cosmic origins is not considered in Einstein's steady –state model; this may be another indication that the manuscript was written early in1931.
[31] Hoyle was also unconvinced by Gamow's postulate of nucelosynthesis in the infant universe and concerned about the problem of the formation of galaxies in an expanding universe (Hoyle 1948).
[32] Bondi and Gold took the view that it was not known whether it was appropriate to apply general relativity to the cosmos on the largest scales (Bondi 1952, p146).
[33] Still, later, Hoyle proposed a modified theory known as the 'quasi steady-state universe', in which the steady-state universe is permeated with local little bangs (Hoyle, Burbidge and Narlikar 1993).

## 6. Conclusions

It should come as no great surprise that when confronted with empirical evidence for an expanding universe, Einstein once considered a steady-state model of the cosmos. There is a great deal of evidence that Einstein's philosophical preference was for an unchanging universe, from his tacit assumption of a static universe in 1917[36] to his hostility to the dynamic models of Friedman and Lemaître when they were first suggested (see section 2). Indeed, an expanding cosmos in which the density of matter remains unchanged seems a natural successor to Einstein's static model of 1917, at least from a philosophical point of view.

However, such a steady-state universe demands a continuous creation of matter and, as Einstein discovered in this manuscript, a successful model of the latter process was not possible without some amendment to the field equations. On the other hand, an expanding universe of varying matter density could be described without any amendment to relativity – and indeed without the cosmological constant, as Einstein suggested in his evolving models of 1931 and 1932 (Einstein 1931b; Einstein and de Sitter 1932). Thus it seems very probable that Einstein decided against steady-state solutions because they were more contrived than evolving models of the cosmos. This suggestion fits very well with our view of Einstein's pragmatic approach to cosmology in these years.[37]

It is also possible that Einstein decided against steady-state models on empirical grounds, i.e., on the grounds that there was no observational evidence to support the postulate of a continuous formation of matter from empty space. It is interesting that, when asked to comment on Hoyle's steady-state model many years later, Einstein is reported to have dismissed the theory as "*romantic speculation*" (Michelmore 1962, p253). This criticism is confirmed in a recently-discovered letter written by Einstein in 1952. Einstein seems highly sceptical of Hoyle's model, and in particular of the postulate of a continuous creation of matter from the vacuum: "*Die kosmologischen Spekulationen von Herrn Hoyle, welche eine Entstehung von Atomen aus dem Raum voraussetzen, sind nach meiner Ansicht viel zu wenig begründet, um ernst genommen zu werden*" or "*The cosmological speculations of Mr Hoyle, which presume the creation of atoms from empty space, are in my view much too poorly grounded to be taken seriously*" (Einstein 1952).[38]

---

[34] The principle observations were the discovery that the distribution of galaxies was significantly different in the distant past, and the discovery of the cosmic microwave background. See (Kragh 1996) chaper 7 for a review.
[35] Alternative versions of steady-state models were suggested, but failed to convince the community (Kragh 1996) chapter 7.
[36] It could be argued that this choice was as much philosophical as empirical because there was no guarantee that an expansion on the largest scales would be detectable by astronomy.
[37] We have argued elsewhere that Einstein's removal of the cosmological constant in 1931, followed by his removal of spatial curvature in 1932, suggests an Occam's razor approach to cosmology (O'Raifeartaigh et al. 2014).

As pointed out in section 4, steady-state models of the cosmos were eventually ruled out by astronomical observation. However, the model of this manuscript presents some useful insights into Einstein's cosmology. In the first instance, it is interesting that Einstein retained the cosmological constant in at least one cosmic model he proposed *after* Hubble's observations; it seems that the widely held view[39] that Einstein was happy to banish the cosmological constant at the first sign of evidence for a non-static universe is not entirely accurate. Instead, it appears that Einstein's attraction to an unchanging universe at first outweighed his dislike of the cosmological constant, just as it did in 1917 - he simply found a new role for the term. Second, we note that, when the flaw in Einstein's steady-state model became evident, he quickly turned to evolving models rather than try again with a more sophisticated steady-state theories; this suggests a dislike of overly contrived solutions, as noted elsewhere (O'Raifeartaigh et al. 2014). Third, Einstein's model reminds us that today's view of an evolving cosmos did not occur as a sudden 'paradigm shift' in the wake of Hubble's observations. Instead, physicists explored a plethora of diverse cosmic models for many years, from the possibility of an expansion caused by a continuous annihilation of matter (Tolman 1930a) to one caused by condensation processes (Eddington 1930), from the conjecture that the redshifts of the nebulae represented a loss of energy by photons (Zwicky 1929) to the hypothesis of a steady-state universe.

We note finally that Einstein's attempt at a steady-state model has some relevance to today's cosmology. His association of the cosmological constant with an energy of space ("*by setting the λ-term, space itself is not empty of energy*") finds new relevance in the context of the recent observation of an accelerated expansion and the hypothesis of dark energy.[40] Indeed, many of today's models of dark energy bear echoes of steady-state theory, not least the hypothesis of 'phantom fields' (Singh et al. 2003).

---

[38] It could be argued that this particular criticism was is a little harsh since the rate of matter creation required for a steady-state universe was far below detectable levels (Hoyle 1948).
[39] See for example (Kragh1999) p34, (Nussbaumer and Bieri 2009) p147, (Nussbaumer 2014b).

We note that today's theories of cosmic inflation[41] also contain many elements of steady-state cosmology. For example, the flat, exponentially expanding metric of most inflationary models is precisely that of Einstein and Hoyle above. Indeed, it has been pointed out by Hoyle (Hoyle 1994 p271) and by other scholars (Barrow 2005) that inflationary models are effectively steady-state cosmologies over an extremely limited timespan.[42] Further, it has been argued (Vilenkin 1983; Linde 1986a,b) that the inflationary process inevitably creates the conditions for further inflation in a never-ending cycle. This 'eternal inflation' model raises the possibility that the observed, evolving universe is a local anomaly in a much larger inflationary ensemble that is in a steady state (Barrow 2005), a vista that is not dissimilar to Hoyle's later proposal of a steady-state universe permeated with local 'little bangs' (Hoyle, Burbidge and Narlikar 1993). These points remind us of the relevance of past models of the universe for cosmology today.

**Acknowledgements**

The authors would like to thank the Albert Einstein Foundation of the Hebrew University of Jerusalem for permission to publish the excerpts from Einstein's original manuscript shown in figures 1 and 2. Cormac O'Raifeartaigh thanks the Dublin Institute of Advanced Studies for access to the Collected Papers of Albert Einstein. Simon Mitton thanks St Edmund's College, Cambridge for the provision of research facilities.

---

[40] See (Peebles and Ratra 2003) for a review.
[41] See (Liddle 1999) for a review.
[42] It might be argued that inflationary models do not demand the continuous creation of matter; however, it is possible to construct steady-state models without this process (Barrow 2005)

**Figure 1**

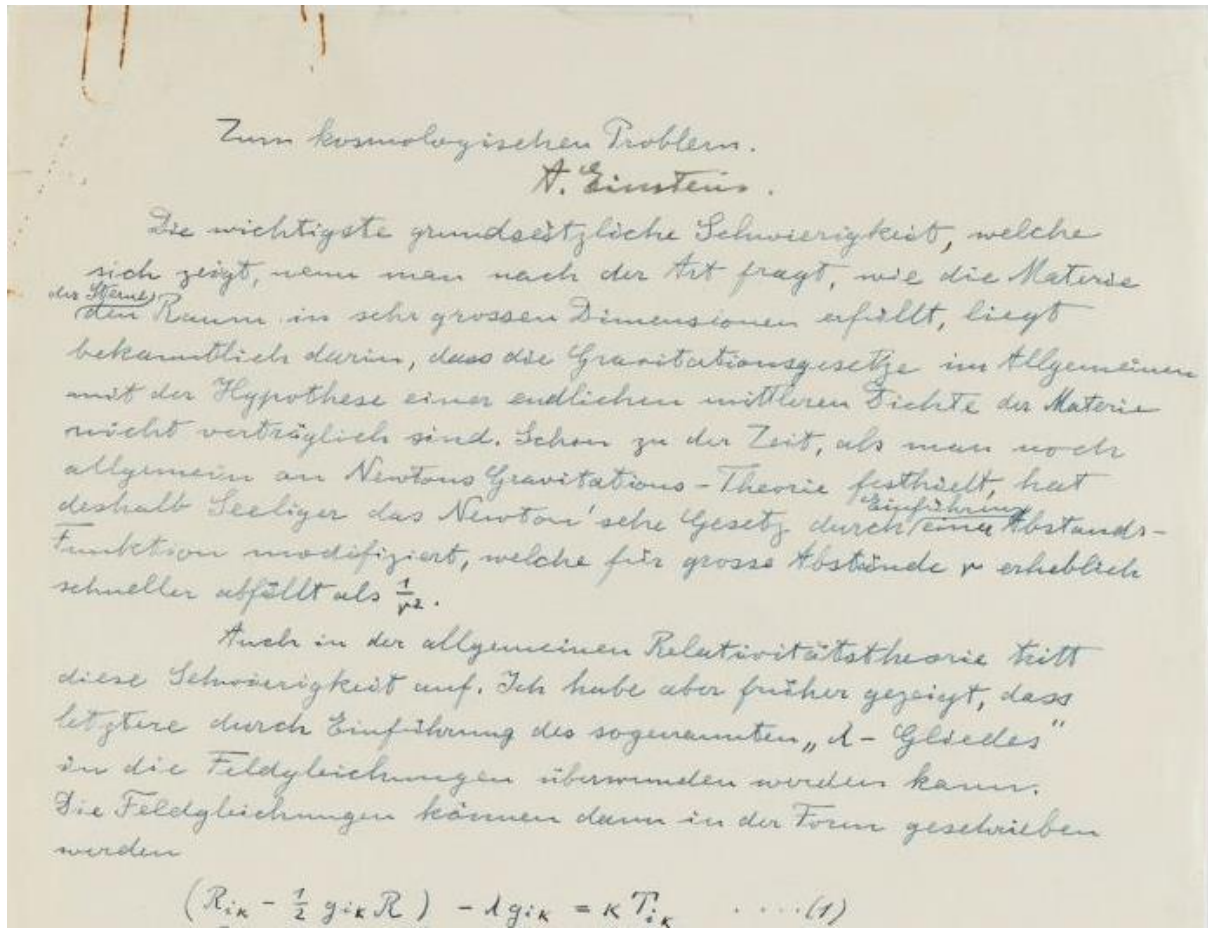

**Figure 1**

An excerpt from the first page of Einstein's steady-state manuscript (Einstein 1931a), reproduced from the Albert Einstein Archive by kind permission of the Hebrew University of Jerusalem

**Figure 2**

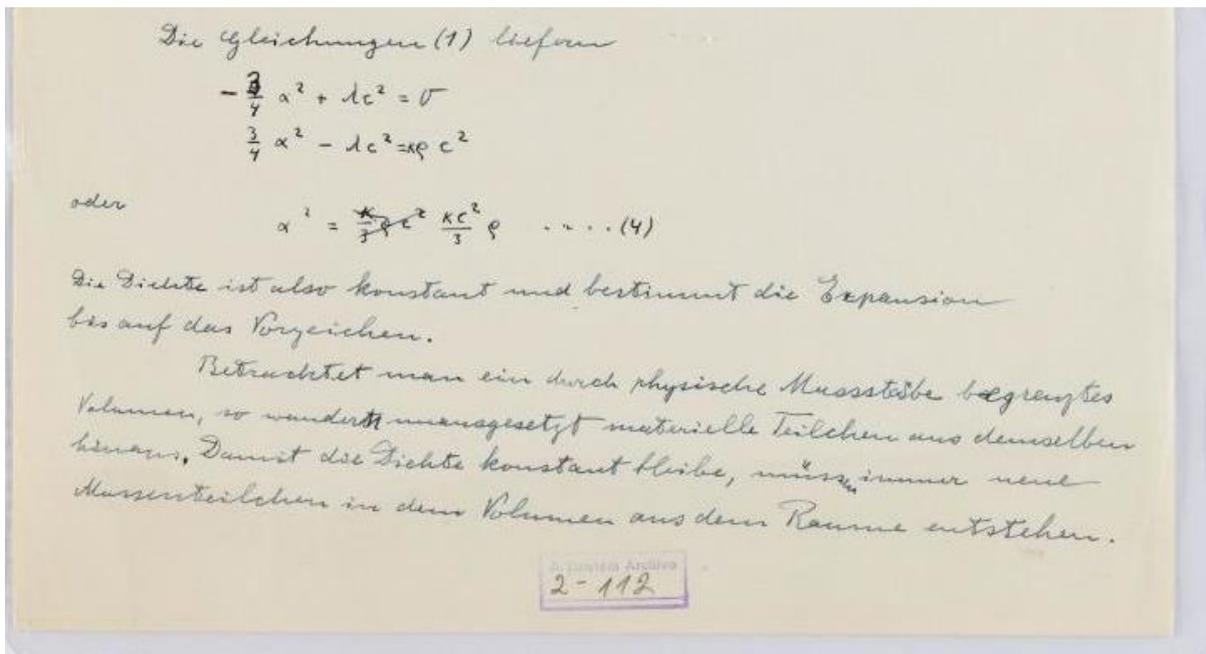

**Figure 2.**

An excerpt from the last page of Einstein's steady-state manuscript (Einstein 1931a), reproduced from the Albert Einstein Archive by kind permission of the Hebrew University of Jerusalem. Equation (4) implies a direct relation between the expansion coefficient $α$ and mean density of matter $ρ$. The sentence immediately below states "*Die Dichte ist also constant und bestimmt die Expansion bis auf das Vorzeichen*" or "The density is therefore constant and determines the expansion apart from its sign". However, the coefficient of $α^2$ in the first of the simultaneous equations was later amended from 9/4 to -3/4, a correction that gives the null result $ρ = 0$ instead of equation (4).